\documentstyle[11pt]{article}
\begin{document}
\title{The Hubbard model on a complete graph: Exact Analytical results.}
\author{Mario Salerno\thanks{Istituto Nazionale per la Fisica della Materia 
and 
Department of Theoretical Physics, University of
Salerno, 84100 Salerno, Italy.
Email:SALERNO@csied.unisa.it}}
\maketitle
\begin{abstract}
We derive the analytical
expression of the ground state of the
Hubbard model with unconstrained hopping at half
filling and for arbitrary lattice sites. 
\end{abstract}
\vskip 4cm
\vskip 2cm PACS: 03.65.Ge, 03.70.+k, 11.10.Lm \newpage
One of the most difficult problems in condensed
matter physics is the characterization of the
spectrum of strongly correlated Fermi systems.
For 1-D integrable systems this can be done by 
Bethe ansatz. For non integrable
systems or for higher dimensional ones, no
generalizations of this method exist, the only
resources being numerical methods.  Exact
diagonalizations of the Hubbard hamiltonian which
take advantage of the conservation of the total
number of electrons $N$ , of the total spin in
the z direction $S_z$ as well as of the
translational invariance of the lattice, has been
performed by several authors \cite{scalapino,parola,seg94}.  
In the case of the
Hubbard model, however, one would like to use as
basis functions the simultaneous eigenfunctions
of $S^2, S_z, N$ and of the operators
characterizing the lattice. Basis functions which are 
invariant under $SU(2), U(1)$ and $S_f$ (permutation group of $f$ 
objects) can be easily constructed in terms of Young 
tableaux filled with 
quanta. By making projections on a 
particular subgroup of the permutation
group $S_f$ one could, in principle, construct 
proper basis functions for the Hubbard
model on arbitrary lattices. Permutation 
group techniques were also used to 
solve the spin $1/2$ Heisenberg chain \cite{Yu93} as well as 
discrete quantum systems of bosons \cite{ms88, se94, pre94}. 
In the present paper we 
consider the case of the Hubbard 
model on a complete graph (unconstrained 
hopping) and show how the above functions 
can lead to an analytical characterization 
of the ground state of the 
system valid for clusters of arbitrary sizes. 
In spite of the simplification introduced by 
the permutational symmetry, the model keeps all the complexity 
of the standard Hubbard model coming from 
the continuous symmetries 
and its solution may be useful for understanding the local 
properties of strongly correlated Fermi systems. 
Moreover, the lacking of exact analytical results 
for the Hubbard 
model in dimensions $D\neq1$ makes the analytical solutions derived 
for this model a good test for numerical computations.

\noindent Let us 
start by introducing the Hubbard Hamiltonian on a complete graph
\begin{equation}
H=-t\sum_{\sigma {i\neq j}}^fc_{i_\sigma
}^{\dagger }c_{j_\sigma }\;{+\;}
U\sum_i^fn_{i_{\uparrow }}n_{i_{\downarrow }} .
\label{hub}
\end{equation}
where $c_{i_\sigma}^{\dagger },c_{j_\sigma }$ , 
$(\sigma =\uparrow$ or $\downarrow )$ 
are standard fermionic creation and annihilation 
operators. The unconstrained hopping in 
Eq.(\ref{hub}) makes the system invariant 
under the action of the permutation group $S_f$. 
In addition to this discrete symmetry there are the 
known symmetries of the Hubbard model i.e. the 
$U(1)$ conservation of the total number of electrons
$N=N_{\uparrow}+N_{\downarrow} = 
\sum_j\left(n_{j_{\uparrow }}+n_{j_{\downarrow}}\right)$, 
the $SU(2)$ invariance under rotations of the total
spin $\stackrel{\rightarrow }{S}$ of the system,
and the particle-hole symmetry. To define the
Hilbert space of the states we denote with
$|3>, |2>,|1>,|0>$, respectively, the doubly
occupied state, the single occupied states with
spin up and spin down, and the vacuum. The states
of the system for a lattice of f sites are 
products of single site states $|n_1n_2...n_f>$ 
with $n_i\in \{0,1,2,3\}$, the dimension of 
the Hilbert space being just $4^{f}$. The
conservation of the number operator allows to
decompose the Hilbert space into a direct sum of
eigenspaces corresponding to a fixed value of
$N$. On the other hand, the invariance of the Hamiltonian
under $S_f$ allows to further block diagonalize
this matrix with respect to the irreps of this
group. Since the total spin of the system
commutes with the permutation operator one can
construct the irreps of $SU(2)$ in terms of Young 
tableaux filled with quanta.
To this end it is convenient to introduce the
quantum number
\begin{equation}
m=N + N_{\uparrow} \label{m}
\end{equation}
and consider all possible partitions
$(m_1,m_2,...,m_f)$ of $m$ into $f$ parts compatible 
with $N, N_{\uparrow}$, with $m_i=0,1,2,3$ and 
$m_1 \geq m_2 \geq ...\geq m_f$ (in eq. (\ref{m}) 
we restrict to $N_{\uparrow }
\geq N_{\downarrow }$ since the other cases
follow from up-down symmetry). We remark that each
partition is associated with an eigenstate 
$|m_1,m_2,...,m_f>$ of $S_z$. This obviously leads to a
family of states (eigenmanifolds of $S_z$)
organized into levels. 
Each level of the hierarchy is
characterized by a specific value of the number
of "particles" $m$ entering the states and by a specific value 
of $S_z=m-3N/2$.  From the above family of
states one constructs eigenmanifolds of $S^2, S_z$
with a given $S_f$ symmetry by filling the quanta
$m_i$, characterizing the states in each level,
in the boxes of a Young tableau according to the
following rules.

\noindent i) The quanta must be not increasing 
when moving from left to
right in each row or when moving down each column
of a given tableau.

\noindent ii) The quanta referring to spin 
up and spin down states ($m_i=1,2$) 
cannot appear more than once in a row.

\noindent iii) The quanta referring to doubly 
occupied states or to empty
states ($m_i=3,0$) cannot appear more than once
in a column.

\noindent These rules directly follow 
from the permutational properties of
the states $|3>,|2>,|1>,|0>$ and from the
symmetry property of Young tableaux. 
To characterize the eigenmanifolds of $S^2, S_z$
with $S=S_z$ (highest spin vectors), one must identify the filled
tableaux corresponding to the highest vectors of
$SU(2)$. This can be done by noting that the 
application of the rising operator on a state associated to
a given filled tableaux produces a state with the same $S_f$ 
symmetry but with a 1 changed into a 2 (1-2 flip) i.e. a state 
associated to a tableaux of the $m+1$ level. We
have therefore, that the tableaux that
satisfy the filling rules also after a 1-2 flip,
are the ones for which $S>S_z$. It is worth to
remark that by a $1-2$ flip two different $m$
tableaux can be in general associated to the same
$m+1$ tableau. In this case one must take linear
combinations $\phi_{\pm}=\psi_1\pm\psi_2$ of the
states corresponding to the $m$ filled tableaux,
to get one state with $S=S_z$ ($\phi_{+}$) and
one with $S=S_z+1$ ($\phi_{-}$). 
By construction, the basis 
functions so obtained are simultaneous
eigenfunctions of $N, S^2, S_z$ which span the
irreps of $S_f$. With respect to these functions
the hamiltonian matrix has a block diagonal form
with the dimension of the blocks given by the
number of "highest weight" tableaux one can fill
with $M$ quanta. The degeneracy of the
corresponding eigenvalues is just $2S+1$ times 
the dimension of the corresponding $S_f$ representation. 
We remark that the above set of rules can be easily implemented on 
a computer using symbolic programs like Mathematica \cite{wo91}. 
We have used a Mathematica implementation of our 
method to study clusters of several sizes. From these studies 
the following conjecture about the ground state of the system can 
be made:
\vskip .1cm
\noindent {\it{Conjecture}}: For $t,U>0$ and for $N<3$ 
the ground state of the system  is always
of symmetry type $\{f\}$. For $f\leq N$ and for $N\geq 3$ 
the symmetry of the ground 
state is of type $\{a_1,a_2,...,a_{r}\}$ with $r$ denoting the 
integer part of $(N+1)/2$, $a_1=f-(N-2)$, $a_i=2$ 
for $i=2,...,r-1$ and $a_r$ equal to 2 or to 1 depending if 
$N$ is even or odd. 
Furthermore, in the ground state S has always
its minimal value ($S=0$ for N even and $S=1/2$ for N odd).
\vskip .1cm
\noindent This conjecture appears natural since in the ground 
state  one expects that the electrons will arrange (for $N< f$) 
in a configuration which minimizes the number of doubly occupied 
states and the tableaux listed above are just the ones which allow this.  
Using this conjecture we can easily derive analytical results for 
the ground state of the system. In particular, at half-filling 
and $f$ even, the ground state is
associated to an $S=0$ ($m=\frac 32f$) tableau of type
$\{2,2,...,2\}$. From 
the filling rules it follows that there are only 
four possible ways of filling a tableau of this type 
with $\frac 32f$ quanta   
i.e. the ones reported in Fig.1. One easily
check that only three of the states associated to
these tableaux are $S=0$ states.   By denoting with 
$\psi_1,\psi _2,\psi _3,\psi _4$ the states associated
with the tableaux from left to right of Fig. 1,
we have that the states $\psi _1,\psi _2$ and
$\psi _3+\psi _4$ are $S=0$ singlets (note that
they are automatically orthogonal) while 
$\psi_3-\psi _4$ is a $S=1$ triplet. From the $S=0$
states one can construct the $3\times 3$
block characterizing the ground state of the
system at half filling for arbitrary f as

\begin{equation}
\left( \matrix{ 2\,U & {\it t} \sqrt{f(f/2 - 1)} & 0 \cr 
* & U & {\it t}\sqrt{f(f/2 + 1)} \cr * & * & 0 \cr }\right)
\label{gsblock}
\end{equation}
(the under diagonal elements were omitted since
the matrix is symmetric),  from which the
ground state energy is readily obtained as
\begin{equation}
E_0=U-\frac{\left( {f^2}\,{t^2}+{U^2}\right)}
{\left({3\,Z}\right)^{\frac 13}} - 
\left({\frac{Z}{9}}\right)^{\frac 13}
\label{gs}
\end{equation}
with
\begin{equation}
Z=9\,f{t^2} U - \sqrt{9\,{f^2}{t^2}{U^2}\left(
9 {t^2}-{f^2}{t^2}-{U^2}\right)
- 3\,\left( {U^6}+{f^6}{t^6}\right)}.\nonumber  
\end{equation}

\noindent 
We see that in both limits $U\rightarrow 0$ and $f\rightarrow 
\infty$,  Eq.(\ref{gs}) reproduces the known
expression $E_0=-t\,f$ of the ground state energy \cite{donvol}.
Furthermore, the degeneracy of $E_0$ is just the
dimension of the $
\{2,2,...,2\}$ representation i.e. 
\begin{equation}
\frac{(f)!}{({\frac f2}+1)!{\frac f2}!}.  \label{gsdim}
\end{equation}
It is worth to note that although the 
lattice is not bipartite, the
ground state at half filling is a singlet 
just as for bipartite lattices.
This seems to indicate that this property is 
true for arbitrary lattices. 
We also remark that the above approach 
can be used to compute the ground state 
for arbitrary fillings. Thus, for 
example, for $N=2$ the ground state energy is found by diagonalizing 
the $2*2$ matrix corresponding to the symmetric states with S=0, 
this giving
\begin{equation}
E_0= 2\,t-f\,t+U/2 - (f^2 t^2 -2 t U + f t U + (U/2)^2)^{1/2},
\end{equation}  
while for $M=3$ is obtained by diagonalizing the 3*3 block 
given by
\begin{equation}
\left( \matrix{U\,- (f/2 -3){\it t}  & -{\it t}/2 \sqrt{f(f - 2)} 
& 1/2 \sqrt{2\,f} \cr 
* & U\,-\,f/2 & -3{\it t}\sqrt{f/2 - 1} 
\cr * & * & -2(f\,-\,3){\it t} \cr }\right).
\label{m3block}
\end{equation}
It is remarkable that for arbitrary fillings $N<f$ 
the ground state energy will 
always be obtained by diagonalizing a matrix of rank less or equal 
to four.

\noindent 
To check our result  we have compared Eq.(\ref{gs}) with numerical
diagonalizations of $H$ up to clusters 
of 16 sites (this is the limit of computability on a supercomputer 
for this model). Thus, for example,  for $f=16,\,t=1,\,U=4$, 
one obtains with  Lanczos method that the ground state energy is 
$E_0=-12.719$ \cite{fanort} while from Eq. (\ref{gs}) 
we get $E_0=-12.7229$. The
difference between these two values represents the 
error involved in the numerical computations at large
values of f. 

\vskip .5cm 
\noindent{\bf {Acknowledgments}} 
\vskip .5cm \noindent Financial
support from the INFM (Istituto Nazionale di
Fisica della Materia), sezione di Salerno, is
acknowledged.

\newpage

\noindent
{\bf {Figure Captions}} 
\noindent Fig.1 
Young tableaux of type $\{2,2,...\}$ filled with 
$M=\frac 32f$ quanta at $N=f$. The rectangular box
represents $\frac f2-2$ identical boxes.

\newpage
\vskip 2cm \centerline{Fig.4} \vskip 1cm 
\[
\begin{picture}(300,300)
\put(30,100){\framebox(15,15){3}}\put(45,100){\framebox(15,15){3}}
\put(30,60){\framebox(15,40){2}}\put(45,60){\framebox(15,40){1}}
\put(30,45){\framebox(15,15){0}}\put(45,45){\framebox(15,15){0}}
\put(90,100){\framebox(15,15){2}}\put(105,100){\framebox(15,15){1}}
\put(90,60){\framebox(15,40){2}}\put(105,60){\framebox(15,40){1}}
\put(90,45){\framebox(15,15){2}}\put(105,45){\framebox(15,15){1}}
\put(150,100){\framebox(15,15){3}}\put(165,100){\framebox(15,15){2}}
\put(150,60){\framebox(15,40){2}}\put(165,60){\framebox(15,40){1}}
\put(150,45){\framebox(15,15){1}}\put(165,45){\framebox(15,15){0}}
\put(210,100){\framebox(15,15){3}}\put(225,100){\framebox(15,15){1}}
\put(210,60){\framebox(15,40){2}}\put(225,60){\framebox(15,40){1}}
\put(210,45){\framebox(15,15){2}}\put(225,45){\framebox(15,15){0}}
\end{picture}
\]

\end{document}